\documentstyle[12pt]{article}
\title{The rare K-decays in the Multiscale Walking Technicolor Model}
\author{ Zhenjun Xiao$^{(1,2)}$
\thanks{E-mail address: dphnu@public.zz.ha.cn}, 
Linxia L\"u$^{2}$, Hongkai Guo$^{2}$ and Gongru Lu$^{1,2}$ \\  
{\small 1. CCAST(World Laboratory) P.O. Box 8730, Beijing 100080, 
P.R.China} \\
{\small 2. Department of Physics, Henan Normal University,
Xinxiang, 453002 P.R.China.} 
\\ }

\begin{document}
\maketitle
\begin{abstract}
We calculate the one-loop $Z^0$-penguin contributions to the rare K-decays,  
 $K^+ \to \pi^+ \nu \bar \nu$, 
$K_L \to \pi^0 \nu \bar \nu$ and  $K_L \to \mu^+ \mu^-$,  
from the  unit-charged technipions $\pi_1$ and 
$\pi_8$ in the framework of the Multiscale Walking Technicolor Model. 
We find that: 
(a) the $\pi_1$ and $\pi_8$ can provide  one  to two orders 
enhancements to the branching ratios of the rare K-decays;
(b) by comparing the experimental data of $Br(K^+ \to \pi^+ \nu \bar \nu)$ 
with the theoretical prediction one finds the lower mass bounds, 
$m_{p8} \geq 249 GeV$ for $F_Q=40GeV$ and $m_{p1}=100 GeV$;
(c) by comparing the experimental data of $Br(K_L \to \mu^+ \mu^-)$ with the 
theoretical predictions one finds the lower bounds on $m_{p1}$ and 
$m_{p8}$,  $m_{p1} \geq 210 GeV$ if only the contribution from 
$\pi_1$ is taken into account, and  $m_{p8} \geq 580 GeV$ for 
$m_{p1}=210 GeV$, assuming $F_Q=40GeV$; 
(d) the assumed ranges of the masses $m_{p1}$ and  $m_{p8}$ in the 
Multiscale Walking Technicolor Model are excluded by the rare K-decay data. 
\end{abstract}

\vspace{0.3cm}


\vspace{0.3cm}

\newcommand{\beq}{\begin{eqnarray}}
\newcommand{\eeq}{\end{eqnarray}}

\newcommand{\paa}{\pi_1}
\newcommand{\pbb}{\pi_8}
\newcommand{\pap}{\pi_1^+}
\newcommand{\pam}{\pi_1^-}
\newcommand{\pbp}{\pi_8^+}
\newcommand{\pbm}{\pi_8^-}
\newcommand{\dsz}{d\overline{s}Z}

\newcommand{\ka}{K^+ \to \pi^+ \nu \overline{\nu}}
\newcommand{\kb}{K_L \to \pi^0 \nu \overline{\nu}}
\newcommand{\kc}{K_L \to \mu^+ \mu^-}
\newcommand{\kcsd}{(K_L \to \mu^+ \mu^-)_{SD}}

\newcommand{\ssa}{\sin^2\theta_W}
\newcommand{\cca}{\cos^2\theta_W}
\newcommand{\mpa}{m_{p1}}
\newcommand{\mpb}{m_{p8}}

\newcommand{\xzxt}{X_0(x_t)}
\newcommand{\cztc}{C_0^{New}}
\newcommand{\brka}{Br(K^+ \to \pi^+ \nu \bar \nu)}
\newcommand{\brkb}{Br(K_L \to \pi^0 \nu \bar \nu)}
\newcommand{\brkc}{Br(K_L \to \mu^+  \mu^-)}
\newcommand{\brkcsd}{Br(K_L \to \mu^+  \mu^-)_{SD}}
\newcommand{\brkcld}{Br(K_L \to \mu^+  \mu^-)_{LD}}


\section*{1. Introduction}

In the framework of the Standard Model(SM), the rare K-decays  
$\ka$, $\kb$ and $\kc$ are all loop-induced semileptonic flavour-changing 
neutral current(FCNC) processes determined by $Z^0$-penguin 
and W-box diagrams. These decay modes have very similar structure, and 
depend on one or two basic functions out of the set ($X(x_t)$, $P_0(X)$, 
$Y(x_t)$, and $P_0(Y)$) \cite{buras974,buras961}.  
The decay $K^+ \to \pi^+ \nu \bar \nu$ is CP conserving and receives 
contributions from both internal top 
and charm quark exchanges, while the decay $K_L \to \pi^0 \nu \bar \nu$ 
proceeds almost entirely through direct CP violation and is completely 
determined by short-distance loop diagrams with top quark 
exchanges. For the 
decay $\kc$, the short-distance part can  also be calculated reliably.

The rare K-decay processes are very good place to probe the effects of 
new physics beyond the SM because these rare decay modes are very clean 
theoretically. Firstly, the short-distance contributions to the rare 
K-decays  can be calculated reliably and the long-distance 
parts to the first 
two decays $\ka$ and $\kb$ are negligibly small. For the decay 
$\kc$, the long-distance contributions from the two-photon 
intermediate state
which are large and difficult to be calculated reliably
\cite{eeg96} because of its unperturbitive nature, 
but we here only consider the new physics effects on the
short-distance contributions to the rare K-decay modes. Secondly, 
the inclusion of next-to-leading order (NLO) QCD corrections reduces 
considerably  the theoretical uncertainty due to the choice of the 
renormalization scales $\mu_t$ and $\mu_c$. Thirdly, the discovery 
of top quark and the measurement of its mass  
reduce significantly another major source of theoretical uncertainty. 
These clean semileptonic decays are also very well 
suited for the determination of CKM matrix elements $V_{ts}$, $V_{td}$
as well as the Wolfenstein parameters $\rho$ and $\eta$, 
but we do not study such topics in this paper. 

As is well-known, Technicolor (TC) \cite{weinberg76} is one of the 
important candidates for the mechanism of naturally breaking electroweak 
symmetry. To generate ordinary fermion masses, extended technicolor (ETC)
\cite{susskind79} models have been proposed. The original ETC models suffer 
from two serious problems: predicting too large flavor changing neutral 
currents (FCNC's) and too small masses for the second and third generation 
fermions. In walking technicolor theories\cite{holdom81}, the first large 
FCNC problem can be resolved and the fermion masses can be increased 
significantly by the large enhancement due to the walking effects of 
$\alpha_{TC}$\cite{holdom81}.  The often-discussed QCD-like one generation 
technicolor model(OGTM) 
\cite{farhi} predicted a rather large oblique 
correction parameter $S$ \cite{peskin92}: $S \approx 1.6$ for $N_{TC}=4$, 
which is contradict with the fitted value of $S= -0.16 \pm 0.14$
\cite{epjc98}. But we know that these estimates do not apply to models 
of walking technicolor because the integrals of weak-current spectral 
functions and their moments converge much more  slowly than they do in QCD 
and consequently simple dominance of spectral integrals by a few resonances 
cannot be correct \cite{lane96}. According to the estimations done in 
refs.\cite{luty93}, the $S$ parameter  can be small or even negative in  
the walking technicolor models\cite{luty93}. To explain the large 
hierarchy of the quark masses, multiscale walking technicolor 
models (MWTCM) are further 
constructed\cite{lane91}. The MWTCM also predicted a large number 
of technirhos and technipions which are shown to be testable in 
experiments\cite{lane96,eichten94}. So it is interesting to study 
the possible contributions to the rare K-decays from the unit-charged 
color-singlet and color-octet technipions in the framework of 
the MWTCM\cite{lane91}. 

In the ``Penguin Box expansion'' (PBE) approach\cite{buras91},  
the decay amplitude for a given decay mode can be written as 
\beq
A(decay) = P_0(decay) + \sum_r P_r(decay)\,F_r(x_t)
\eeq
where the $F_r(x_t)$ are the basic, universal, process independent but 
$m_t$-dependent functions \footnote{ The complete set of functions 
$F_r(x_t)$ include:   
$S_0(x_t)$, $X_0(x_t)$, $Y_0(x_t)$, $Z_0(x_t)$, $E_0(x_t)$, 
$D_0^{'}(x_t)$ and $E_0^{'}(x_t)$,  as given explicitly in 
ref.\cite{buras974}} with corresponding coefficients $P_r$ characteristic 
for the decay under consideration; and the $m_t$-independent term $P_0$ 
summarizes contributions stemming from internal quarks other than the top, 
in particular the charm quark.

In a previous paper\cite{ka982}, we calculated the $Z^0-$penguin and box 
contributions from the unit-charged technipions to the rare K-decays 
in the framework of the one generation technicolor model\cite{farhi}.
In this paper, we will estimate the corresponding contributions to the 
rare K-decays in the framework of MWTCM\cite{lane91}. 
Our strategy for the current work is rather simple: we evaluate the
$Z^0$-penguin and box diagrams induced by the charged technipions, 
compare the relevant analytical expressions of effective couplings with
the corresponding expressions in the SM, separate the new functions 
$C_0^{New}$ and $C_{NL}^{New}$,  which summarize the effects of the 
new physics beyond the SM, and 
finally combine the new functions with their counterparts in the SM and use 
the new basic functions  directly in the calculation for specific decays. 

From the  numerical calculations, we find that 
the unit-charged color-singlet and color-octet technipions $\paa$ and 
$\pbb$  appeared in the MWTCM can provide  two to 
three orders enhancement 
to the branching ratios of the rare K-decays. 
The contribution from the color-singlet $\paa$ is
positive but relatively small in size, and therefore 
the color-octet technipion $\pbb$ dominates the total contributions.  

By comparing the experimental data of the rare K-decays with  the 
theoretical predictions one can obtain the lower bounds on the masses 
of charged technipions\footnote{In ref.\cite{lane91}, the authors used 
the symbol $\pi_{\bar D U}$ and $\pi_{D \bar U}$ to denote the unit-charged 
color-octet technipions, we here use the symbol 
$\pbb$. We also use the $\paa$  to denote the physical mixed state of 
the $P_1^+$ and $P_2^+$, and use the $\mpa$ and $\mpb$ to denote the 
masses of $\paa$ and $\pbb$. }. For the decay mode $\ka$, the lower mass 
bounds are $\mpb \geq 249, 228GeV$ 
for $F_Q=40GeV$ and $\mpa=100, 200GeV$ respectively;
For the decay mode $\kc$, the lower mass bound on $\mpa$ is 
$\mpa \geq 210 GeV$ if only the contribution from 
$\paa$ is taken into account and assuming $F_Q=40GeV$, while 
the lower mass bound on $\mpb$ is $\mpb \geq 580 GeV$ 
assuming  $F_Q=40GeV$ and $\mpa=210 GeV$. For $F_Q=30GeV$, the above 
lower mass bounds will be increased by about 50 GeV. 
For $F_Q=40GeV$ and $\mpb=490 GeV$ the whole parameter space for 
$\mpa$ is excluded completely. 
For the decay mode $\kb$, however, no lower mass bounds could be 
derived because of the low sensitivity of the corresponding
experimental data. The assumed ranges of the masses $\mpa$ and 
 $\mpb$ in the Multiscale 
Walking Technicolor Model\cite{lane91} are excluded by the rare K-decay 
data, and therefore the specific model\cite{lane91} itself is disfavored 
by the data.

This paper is organized as follows. In Sec.2 we describe the basic 
structures of the MWTCM, briefly review  
the properties of the charged technipions $\paa$ and $\pbb$, and present 
the effective $Z^0$-penguin couplings with the inclusion of technipion 
contributions. In the following three sections, we calculate the new 
contributions to the decays $\ka$, $\kb$ and $\kc$ respectively and try to  
extract the possible lower mass bounds on charged technipions by 
comparing the theoretical predictions with the corresponding experimental
data. The conclusions and discussions are included in the final section.

\section*{2. Effective $\dsz$ coupling and relevant formulae}

The rare FCNC K- and B-decays have been 
investigated at the NLO level within the framework of the SM. Consequently, 
the relevant formulae and the systematic analysis in the SM can be found 
easily in new review papers \cite{buras974,buras961}. The impact of some 
scenarios of new physics on the rare K- and B-decays has been considered 
for instance in refs.\cite{ka982,bigi91,misiak97,burdman98,lu96}. In this 
paper we will 
investigate the contributions to the rare FCNC K-decays from the 
unit-charged technipions $\paa$ and $\pbb$ in the framework of the MWTCM.

\subsection*{2.1 Basic structures of the MWTCM}

In ref.\cite{lane91}, the authors constructed a specific multiscale walking 
technicolor model and investigated its phenomenology. 
The major features of this model relevant with our 
studies are the following:
\begin{enumerate}
\item
This model contains one doublet $\psi=(\psi_U, \psi_D)$ of color-singlet
technifermions in the antisymmetric tensor representation $A_2$ of 
$SU(N_{TC})$; one doublet of color-triplet techniquarks, $Q= (U, D)$; and 
$N_L$ doublets of color-singlet technileptons, $L_m=(N_m, E_m)$, $m=1, 
\cdots, N_L$.
Under the gauge group $SU(N_{TC}) \otimes SU(3) \otimes SU(N_L)
\otimes SU(2)_I$\footnote{Which is obtained by two steps of breaking from
the ETC gauge group $SU(N_{ETC})_1 \otimes SU(N_{ETC})_2$ as described in 
ref.\cite{lane91}.} the technifermions are 
\beq
T_{3L,R} \equiv \psi_{L,R} \in (A_2, 1, 1, 2), \nonumber\\
T_{2L,R} \equiv Q_{L,R} \in (N_{TC}, 3, 1, 2), \\
T_{1L,R} \equiv L_{L,R} \in (N_{TC}, 1, N_L, 2). \nonumber
\eeq

\item
They assumed that the technifermion chiral-symmetry breaking scales 
$\Lambda_i$, the condensates 
$<\bar T_i T_i>$, the $\pi_T$ decay constant $F_i$ ($i=L, Q, \pi$) may 
be estimated from the corresponding QCD parameters by naive scaling and 
large $N_{TC}$ arguments.  They studied the phenomenology under 
the limits of $\Lambda_{\pi} >> \Lambda_Q \cong \Lambda_L$ and 
$F_{\psi} >> F_Q \cong F_L$ with the constraint
\beq
\sqrt{N_L F_L^2 + 3F_Q^2 + F_{\psi}^2} = v =246GeV
\eeq
where $F_Q=20 \sim 40 GeV$.
\item
This  model predicted a rich spectrum of technipions. Among them are 
unit-charged color-octets $\pi^a_{\bar D U}$ and $\pi^a_{\bar U D}$,  
and unit-charged color-singlets $P_1^+$ and $P_2^+$, which will contribute 
to the rare K-decays in question through the $Z^0-$penguin and box diagrams. 

\item
The authors calculated the dijet and technipion production rates at 
$\bar p p$ colliders by using two sets of input parameters. The Set-A 
and Set-B mass parameters (all masses are in GEV) are :
\beq
Set-A: \ \ \ F_L=28, \; F_Q=29, \;  
M_{P_1^+}=172, \; M_{P_2^+}=251, \; 
M_{\pi_{\bar D U}}=261, \cdots, \label{seta}
\eeq
and 
\beq
Set-B: \ \ \ F_L=41,\; F_Q=43, \;   
M_{P_1^+}=218, \; M_{P_2^+}=311, \; 
M_{\pi_{\bar D U}}=318, \cdots, \label{setb}
\eeq
\end{enumerate}

The technipion $\pi^a_{\bar DU}$ in ref.\cite{lane91} is just
the same technipion as the $P_8^{+}$ appeared in the one-generation 
technicolor model\cite{farhi,eichten86}, and the technipions 
$P_1^+$ and $P_2^+$ are mainly $\bar E N$ and $\bar D U$ with small 
$\overline{\psi_D} \psi_U$ piece and therefore the mixed state 
$\paa$ of the 
$P_1^+$ and $P_2^+$ is  the same kind of technipion as $P^+$ given in 
refs.\cite{farhi,eichten86}.  We will study the new contributions to 
the rare K-decays from the physical mixed state $\paa$ instead of the two 
technipions $P_1^+$ and $P_2^+$, for the sake of simplicity.

If these technipions are relatively light as assumed in ref.\cite{lane91} 
they  will contribute to various production and decay processes effectively. 
At the Tevatron and LHC, they can be pair produced copiously, as discussed 
systematically in refs.\cite{lane96,lane91,eichten94,john96,lane97}.
In this paper, we calculate the new  
contributions to the rare K-decays  from the  
$\paa$ and $\pbb$ as described  in the MWTCM \cite{lane91}.
For the rare K-decays under consideration, 
the charged technipions may contribute through the $\dsz$-penguin and box 
diagrams by effective technipion-fermion pair
and $Z^0$-technipion pair  Yukawa couplings.

The color-singlet technipion $\paa$ is the closest  analog to the charged 
Higgs boson $H^{\pm}$ in the two Higgs doublet model\cite{2hdm}, 
but the color-octet technipion is rather different with the $H^{\pm}$
since it carries color and therefore is involved in the QCD and ETC strong 
interaction as well as the electroweak interactions.  It is this 
fact which makes the difference between the charged Higgs bosons 
and unit-charged technipions.

The most model-independent limit on the mass of $H^{\pm}$\cite{martin96}, 
$M(H^{\pm}) \geq 44 GeV$, also apply to $\paa$. The color-octet technipion
$\pbb$ receives QCD, electroweak and extended technicolor (ETC)  
contributions to its mass,  one previous 
estimation predicted that $M(\pbb) \approx 200 GeV$\cite{eichten86}. 
In walking technicolor, however, the large ratio $< \bar T T >_{ETC}/
< \bar T T >_{TC}$ will enhance technipion masses, and consequently the 
technipions in walking technicolor models are generally heavier than those 
in the ordinary OGTM. Unfortunately, it is almost impossible to predict the 
masses of technipions reliably at present. What one can do is 
a qualitative estimation about the range of those masses, 
as has been done in ref.\cite{lane91}, where the authors estimated the 
contributions to technipion masses from different sources and gave the
typical ranges: $M(P^+)= 170 \sim 320 GeV$ and 
$M(\pi_{\bar D U})=250 \sim 320GeV$ corresponding their choice for 
different sets of parameters. 
In this paper, we treat the masses of $\paa$ and $\pbb$ 
as semi-free parameters, 
varying in the ranges of $50 GeV \leq \mpa \leq 400 GeV$ 
and $100 GeV \leq \mpb \leq 600 GeV$ respectively. Generally speaking, 
the color-singlet $\paa$ should be lighter than the color-octet $\pbb$. 

The ETC interaction couples technifermions to quarks and leptons, and so 
governs the couplings between technipions and fermion pairs. Such 
effective Yukawa couplings are therefore ETC model dependent. According 
to the conventional wisdom, which is inspired by analogy with the
SM, the technipions couple essentially to fermion masses. In other words, 
these effective Yukawa couplings are Higgs-like, and so the couplings 
between technipions and heavy fermions (especially the third generation 
fermions) will be dominant. According to the estimations done by 
J. Ellis et al. 
\cite{eichten86,ellis81}, the effective Yukawa couplings of   
charged technipions to fermion pairs can be 
written as \cite{eichten86,ellis81},
\beq
(\frac{-i}{F_{\pi}})\, \pap \, \left \{ V_{KM}
\left( m_d\,\overline{u_L}\,d_R
-m_u\, \overline{u_R}\,d_L \right) \sqrt{2/3} 
-\sqrt{6}m_e\,\overline{\nu_{eL}}\,e_R \right \} + h.c. \\
\label{p1ff}
(\frac{-i}{F_{\pi}})\, \pi_{8\alpha}^+\, \left \{ 
V_{KM} 
\left(m_d \overline{u_L}\, \lambda^{\alpha}\, d_R -
m_u\overline{u_R}\,\lambda^{\alpha}\,d_L 
\right)  \right \}\, 2 + h.c, 
\label{p8ff}\hspace{2.6cm}
\eeq
where the $L, R = ( 1 \mp \gamma_5)/2$, the $u$ and $d$ stand for the 
up and down type quarks $(u, c, t)$ and $(d, s, b)$ respectively, 
the $e$ denotes the leptons $(e, \mu, \tau)$, the $\lambda^{\alpha}$
$(\alpha=1, \cdots, 8)$ are the Gell-Mann $SU(3)_C$ matrices,  
 the $V_{KM} $ is the element of CKM matrix. The technipion decay constant 
$F_{\pi}$ is model dependent: $F_{\pi}=123 GeV$ in the often-discussed OGTM, 
while $F_{\pi} = F_Q =20 \sim 40GeV$ in the multiscale walking 
technicolor model\cite{lane91} in order to produce the correct masses for 
the gauge bosons $Z^0$ and $W$. 

The gauge interactions of technipions with the standard 
model gauge bosons occur dynamically through technifermion loops.  
At low energy scales well below the Technicolor scale 
their couplings to gauge bosons can be evaluated reliably by using
well-known techniques of current algebra or effective lagrangian methods
\cite{lane91,eichten86,peskin81}. 
The relevant gauge couplings of $Z^0$ gauge boson to   
charged technipion pairs can be 
written as \cite{eichten86,li84},
\beq
Z \pap \pam : \hspace{1cm}  
-ig\frac{1-2\sin^2\theta_W}{2\cos\theta_W}(p^+ - p^-)\cdot \epsilon
\hspace{4cm} \\
Z \pi_{8\alpha}^+\pi_{8\beta}^-: \hspace{1cm}
 -ig\frac{1-2\sin^2\theta_W}{2\cos\theta_W}(p^+ 
- p^-)\,\delta_{\alpha \beta} \cdot \epsilon \hspace{3.5cm} 
\eeq
where the $\sin\theta_W$ is the Weinberg angle, the $p^+$ and $p^-$ are 
technipion momenta and the $\epsilon$ is the polarization vector 
of $Z^0$ gauge boson.

\subsection*{2.2 The $Z^0$-penguin and box diagrams }

The new one-loop diagrams for the induced $\dsz$ couplings due to 
the exchange of the technipions $\paa$ and $\pbb$ are shown in Fig.1. 
The Fig.2 shows the box diagrams when the W gauge boson internal lines are 
replaced by color-singlet technipion lines. The color-octet $\pbb$ does not 
couple to the $l\nu$ lepton pairs, and therefore does not present in the 
box diagrams.
The corresponding one-loop diagrams in the SM were evaluated long time 
ago and can be found in ref.\cite{inami81}. We here just draw the new 
one-loop diagrams
where the technipion propergators are inserted in all possible ways under 
the t' Hooft-Feynman gauge. 

Because of the lightness of the s and d quarks  when 
compared with the large top quark mass and the technipion masses 
we set $m_s=0$ and $m_d=0$
in the calculation. We will use dimensional regularization to regulate 
all the ultraviolet divergences in the virtual loop corrections 
and adopt the $\overline{MS}$ renormalization scheme.  It is easy to show 
that all ultraviolet divergences are canceled for $\paa$ and 
$\pbb$ respectively,  and therefore the total sum is also finite.

By analytical evaluations of the Feynman diagrams as shown in Fig.1, 
we find the effective $\dsz$ vertex induced by the $\paa$ exchanges,
\beq
\Gamma^I_{Z_{\mu}} = 
\frac{1}{16 \pi^2}\frac{g^3}{\cos\theta_W}\; \sum_{j} \lambda_j\,
\overline{s_L}\, \gamma_{\mu}\, d_L\; C_0^{New}(y_j)
\label{bsza} 
\eeq
with   
\beq
C_0^{New}(y_j) =\eta_{TC}^a \; \left[ \frac{y_j(-1 +2\ssa -3y_j 
+2\ssa y_j)}{8(1-y_j)} - \frac{\cos^2\theta_W y_j^2}{2(1-y_j)^2}\ln[y_j] 
\right]\label{cza} 
\eeq
and 
\beq
\eta_{TC}^a =\frac{\mpa^2}{24\sqrt{2}F_Q^2 G_F M_W^2} 
\eeq
where  $\lambda_j=V_{js}^*V_{jd}$, $y_j=m_j^2/\mpa^2$, and the 
$G_F=1.16639(2)\times 10^{-5} (GeV^{-2})$ is the Fermi coupling constant.

For the case of color-octet $\pbb$, the effective 
$\dsz$ vertex induced by the $\pbb$ exchanges is the form of,
\beq
\Gamma^{II}_{Z_{\mu}} = 
\frac{1}{16 \pi^2}\frac{g^3}{\cos\theta_W}\; \sum_{j} \lambda_j\,
\overline{s_L}\, \gamma_{\mu}\, d_L\; C_0^{New}(z_j)
\label{bszb}
\eeq
with   
\beq
C_0^{New}(z_j)=\eta_{TC}^b \,\left[ 
\frac{z_j(-1 +2\ssa -3z_j +2\ssa z_j)}{8(1-z_j)} -
 \frac{\cos^2\theta_W z_j^2}{2(1-z_j)^2}\ln[z_j] \right]
\label{czb} 
\eeq
and 
\beq
\eta_{TC}^b =\frac{\mpb^2}{3\sqrt{2}F_Q^2 G_F M_W^2} 
\eeq
where $z_j=m_j^2/\mpb^2$.

When compared with  the eq.(2.6) of ref.\cite{inami81}, one can see that 
the  $C_0^{New}(y_j)$ and $C_0^{New}(z_j)$ in eqs.(\ref{cza},\ref{czb}) are  
just the same kind of terms as the 
function $\Gamma_Z$ in eq.(2.7) of ref.\cite{inami81} or the basic function
$C_0(x_i)$ in eq.(2.18) of ref.\cite{buras974}. The $C_0^{New}(y_j)$ 
describes the contributions to the $\dsz$ vertex from the color-singlet 
technipion $\paa$, while the $C_0^{New}(z_j)$ describes the 
contributions from the color-octet technipion $\pbb$. 

In the above calculations, we used the unitary relation $\sum_{j=u,c,t}
\lambda_j\cdot \;constant=0$ wherever possible, and  neglected all terms 
proportional to $p^2$, $p'^2$ and $p\cdot p'$ $(p'= p-k)$. This is a 
conventional approximation. We also used the functions $(B_0, B_{\mu}, 
C_0, C_{\mu}, C_{\mu\nu})$ whenever needed to make the integrations, 
and the
explicit forms of these complicated functions can be found, for instance, 
in the Appendix-A of ref.\cite{cho91}.

For the color-singlet $\paa$, it does couple to $l\nu$ pairs through box 
diagram as  shown in Fig.2, but the 
relevant couplings are strongly suppressed  by the lightness of $m_l$.
Even for the $\tau$ lepton, the corresponding cross section still be 
suppressed by an additional factor of $m_{\tau}^2/F_Q^2 
\sim 10^{-3}$. Consequently, we can neglect  the tiny contributions 
from $\paa$ through the box diagrams safely. The color-octet $\pbb$ does 
not couple to any lepton pairs, and therefore can not contribute to the 
rare K-decays through the Box diagrams. 
In short,  the technipion $\paa$ and $\pbb$ contribute effectively to the
rare K-decays through the $Z^0$-penguin diagrams only. 
We therefore can include the contributions from $\paa$ and $\pbb$ 
to the rare K-decays by simply adding the functions  
$\cztc$ with the function $C_0(x_i)$  given in ref.\cite{buras974}. 

\subsection*{2.3 Basic functions at the NLO level}

Within the standard model, the decay $\ka$ depends on 
the functions $X(x_t)$ and $X_{NL}^l$ relevant for the top part and  
charm part respectively, and the decay $\kb$ depends
on one basic function $X(x_t)$, while the short-distance part of the 
decay $\kc$ depends on the 
functions $Y(x_t)$ (the top part) and $Y_{NL}$ (the charm part)
\cite{buras974,buras961},  
\beq
X(x_t) &=& X_0(x_t) + \frac{\alpha_s}{4\pi} X_1(x_t), \label{xxt} \\
Y(x_t) &=& Y_0(x_t) + \frac{\alpha_s}{4\pi} Y_1(x_t), \label{yxt} \\
X_{NL}^l &=& C_{NL} -4 B_{NL}^{1/2}, \label{xnl}  \\ 
Y_{NL} &=& C_{NL} - B_{NL}^{-1/2}\label{ynl}
\eeq
where the functions $X_0(x_t)$ and $Y_0(x_t)$ are leading contributions 
\beq
X_0(x_t) = C_0(x_t)-4B_0(x_t),  \ \ 
Y_0(x_t) = C_0(x_t)-B_0(x_t)
\eeq
and the functions $X_1(x_t)$ and $Y_1(x_t)$ are QCD corrections, and 
finally the function $C_{NL}$ is the $Z^0$-penguin part in the charm 
sector, the functions $B_{NL}^{1/2}$ 
and $B_{NL}^{-1/2}$ are the box contributions in the charm sector, 
relevant for the case of final state neutrinos (leptons) with weak 
isospin $T_3=1/2$ ($-1/2$) respectively. 
In ref.\cite{buras974}, the authors also defined the 
functions $P_0(X)$ and $P_0(Y)$ for the decay $\ka$ and $\kc$ 
respectively, 
\beq
P_0(X)^{SM} &=& \frac{1}{\lambda} \left[ \frac{2}{3}X_{NL}^e 
+ \frac{1}{3}X_{NL}^{\tau} \right], 
\label{p0x} \\
P_0(Y)^{SM} &=& \frac{Y_{NL}}{\lambda^4}.
\label{p0y} 
\eeq
The $P_0$ functions describe the contributions from the charm sector.

When the new contributions from charged technipions are included, the 
functions $X$, $Y$ and $P_0$ can be written as
\beq
X(x_t, y_t, z_t) &=& X(x_t) + C_0^{New}(y_t) + C_0^{New}(z_t),
\label{xxtt} \\
Y(x_t, y_t, z_t) &=& Y(x_t) + C_0^{New}(y_t) + C_0^{New}(z_t), 
\label{yxtt}\\
P_0(X) &=& P_0(X)^{SM} + 
\frac{1}{\lambda} \left[ C_{NL}(\paa) + C_{NL}(\pbb) \right] 
\label{p0xt} \\
P_0(Y) &=& P_0(Y)^{SM} + \frac{1}{\lambda^4} \left[  
 C_{NL}(\paa) + C_{NL}(\pbb) \right] 
\label{p0yt} 
\eeq
Where the functions $C_0^{New}(y_t)$ and $C_0^{New}(z_t)$ are given in 
eqs.(\ref{cza},\ref{czb}).  
For completeness, we present the expressions for the functions 
$C_0(x_t)$, $B_0(x_t)$, $X_1(x_t)$, $Y_1(x_t)$, $C_{NL}$, $B_{NL}^{1/2}$,  
$B_{NL}^{-1/2}$, $C_{NL}(\paa)$ and $C_{NL}(\pbb)$ in the Appendix. 
One can also find the explicit expressions for the relevant functions 
within the Standard Model in refs.\cite{buras974,buras961}.

In this paper we do not investigate the uncertainties in the prediction
for the branching ratios of rare K-decays 
related to the choice of the renormalization scales 
$\mu_t$ and $\mu_c$ in the top part and the charm part, respectively. 
In the numerical calculations, we  fix the relevant parameters as follows 
and use them as the Standard Input (all masses are in GeV):
\beq
M_W&=&80.2,\;  G_F=1.16639\times 10^{-5} GeV^{-2},\; 
 \alpha=1/129,\; \ssa =0.23, 
\nonumber \\
m_c &\equiv & \overline{m_c}(m_c)=1.3,\; 
m_t \equiv \overline{m_t}(m_t)=170, \;
\mu_c=1.3,\; \mu_t =170, 
\nonumber \\
\Lambda^{(4)}_{\overline{MS}}&=&0.325, \; 
\Lambda^{(5)}_{\overline{MS}}=0.225, \; 
A=0.84, \;  \lambda=0.22, \;  \rho=0,\;  \eta=0.36
\label{sip}
\eeq
where the $A, \lambda, \rho$ and $\eta$ are Wolfenstein parameters at 
the leading order.  For 
$\alpha_s(\mu)$ we use the two-loop expression as given in 
ref.\cite{buras961}, 
\beq
\alpha_s(\mu) = \frac{4\pi}{\beta_0\, \ln(\mu^2/\Lambda^2)}\;
\left[ 1 - \frac{\beta_1}{\beta_0^2}\cdot 
\frac{\ln \ln(\mu^2/\Lambda^2)}{\ln(\mu^2/\Lambda^2)} \right], 
\eeq
with
\beq
\beta_0 = \frac{33-2N_f}{3}, \ \ \beta_1 = 102 -10 N_f -8N_f/3
\eeq
where the $N_f$ is the number of quark flavours.  

In the SM, using the input parameters as given in eq.(\ref{sip}), 
one obtains $X(x_t)=1.539$, $P_0(X)^{SM}=0.352$, $Y(x_t)=1.04$, and 
$P_0(Y)^{SM}=0.150$. These values are in very good agreement 
with those given in ref.\cite{buras974}. 

When the contributions from the technipions are included, the $X$, 
$Y$ and $P$ functions generally  depend on the masses of the $\paa$ and 
$\pbb$. The color-octet $\pbb$ dominate the total contribution because 
of the color enhancement.

Fig.3a shows the $\mpb$ dependence of the $X(x_t)$ function assuming 
$F_Q=40GeV$. The short-dashed line is the contribution in the SM.
The $\paa$ provide a positive contribution, the typical value is 
$ X(x_t)(\paa) = 1.352$ for $ \mpa = 100GeV$. The $\pbb$ can 
provide a rather large positive contribution to the function $X(x_t)$  
when it is light, one typical value is $ X(x_t)(\pbb) = 3.127$ for 
$ \mpb = 300GeV$. The contribution will become negative for $\mpb 
\geq 531 GeV$. The long-dashed line shows the total contribution for both 
$\paa$ and $\pbb$. The solid line represents the total contribution. 

Fig.3b shows the $\mpb$ dependence of the function $Y(x_t)$ assuming 
$F_Q=40GeV$. The short-dashed line is the contribution in the SM.
The charged technipions provide the same kinds of contributions to the 
function $Y(x_t)$ as that to $X(x_t)$. The dot-dashed line shows the 
contribution from the $\paa$ for $\mpa =100 GeV$. The long-dashed line 
shows the contribution from $\pbb$ and  the solid line represents 
the total contribution. For smaller $F_Q$, the size of the functions 
$X(x_t)$ and $Y(x_t)$ will become more larger.

Fig.4a and Fig.4b are plots of functions $P_0(X)$ and $P_0(Y)$ vs 
$\mpb$. The short-dashed line shows the $P_0(X)$ and $P_0(Y)$ in the 
SM, the dot-dashed lines are the contributions from $\paa$ 
for $\mpa=100 GeV$. 
The typical values  are $ P_0(X)(\paa) = P_0(Y)(\paa)=0.012$ for 
$ \mpa = 100GeV$ and $ P_0(X)(\pbb) = P_0(Y)(\pbb)=0.152$ for 
$ \mpb = 200GeV$.  The solid line again shows the total contribution.

\section*{3. The decay $\ka$ }

Within the Standard Model, 
the effective Hamiltonian for $\ka$  are now available at the NLO level
\cite{buras974}, 
\beq
{\cal H}_{eff} = \frac{G_F}{\sqrt{2}}\frac{\alpha}{2\pi \ssa}
\sum_{l=e, \mu, \tau} \left( V^*_{cs}V_{cd} X^l_{NL} +
 V^*_{ts}V_{td} X(x_t) \right) 
(\overline{s}d)_{V-A}(\overline{\nu_l}\nu_l)_{V-A}  
\label{heff}
\eeq
where the functions $X(x_t)$ and $X_{NL}^l$ have been given in 
eqs.(\ref{xxt},\ref{xnl}). 

Using the effective Hamiltonian $(\ref{heff})$ and summing over the three 
neutrino flavors one finds
\beq
Br(\ka) = \kappa_+ \cdot \left[ 
\left( \frac{Im \lambda_t}{\lambda^5} X(x_t) \right)^2 
+\left( \frac{Re \lambda_c}{\lambda}P_0(X)^{SM} + 
\frac{Re \lambda_t}{\lambda^5}X(x_t) \right)^2 \right] \label{brkasm}
\eeq
where $\kappa_+ =4.11 \times 10^{-11} $\cite{buras974}, 
and $\lambda=0.22$ is the Wolfenstein parameter. 

When the new contributions are included, one finds
\beq
Br(\ka) = \kappa_+ \cdot \left[ \left( \frac{Im \lambda_t}{\lambda^5}
X(x_t,y_t,z_t) \right)^2 \right. \hspace{3cm} \nonumber\\
\left. \hspace{5cm}+ \left( \frac{Re \lambda_c}{\lambda}P_0(X) + 
\frac{Re \lambda_t}{\lambda^5} X(x_t,y_t,z_t)\right)^2 \right] 
\label{brkat}
\eeq

Within the SM, using the input parameters of eq.(\ref{sip}), one finds
\beq
\brka = 8.72 \times 10^{-11},\label{brsm}
\eeq
which is consistent with the result given in ref.\cite{buras974}. 
When the contributions due to $\paa$ and $\pbb$ are included, the size of 
the corresponding branching ratios depends on the masses $\mpa$ and 
$\mpb$.   Using the input parameters of eq.(\ref{sip}) and assuming 
$F_Q=40 GeV$, $50 GeV \leq \mpa \leq 400 GeV$ and   $100 GeV \leq \mpb 
\leq 600 GeV$, one finds 
\beq
1.07\times 10^{-10} \leq \brka \leq 3.21 \times 10^{-10}  
\eeq
if only the $\paa$'s contribution is included, and 
\beq
7.34\times 10^{-11} \leq \brka \leq 3.69 \times 10^{-9}  
\eeq
if only the $\pbb$'s contribution is included, and 
\beq
9.14\times 10^{-11} \leq \brka \leq 4.81 \times 10^{-9}  
\eeq
if the $\paa$'s and $\pbb$'s contribution are all included.

For the typical values of $F_Q=40GeV$, $\mpa=200 GeV$ and $\mpb=300 
GeV$, one has 
\beq
\brka = \left \{ 
\begin{array}{ll} 
8.72\times 10^{-11} &  {\rm in \ \ the \ \ SM} \\
1.67\times 10^{-10} & {\rm only}\ \ \paa\ \ {\rm considered} \\
6.33\times 10^{-10} & {\rm only}\ \ \pbb \ \ {\rm considered} \\
8.27\times 10^{-10} & {\rm both}\ \ \paa\ \ {\rm and}\ \ \pbb
\ \  {\rm considered}.
\end{array} \right.
\label{brtc}
\eeq

The new experimental bound on $\brka$ is\cite{adler97}:
\beq
\brka_{exp} = 4.2^{+9.7}_{-3.5} \times 10^{-10}
\label{brkaexp}
\eeq
which is close to the SM expectations (\ref{brsm}), and begins to cross the 
range of the theoretical expectations when the 
new contributions from  the charged 
technipions are included. There is no lower limit on $\mpa$ if we neglect 
the contribution from the $\pbb$. The lower mass bound on $\pbb$ depends on 
the values of the $F_Q$ and $\mpa$:
\beq
\mpb \geq 249,\; 228GeV
\label{massb}
\eeq
for $F_Q=40GeV$ and $\mpa=100, 200GeV$ respectively.  
For $F_Q=30GeV$ and $\mpa=200GeV$ , one has $\mpb \geq 341GeV$.

The Fig.5a shows the $\mpa$ dependence of the branching ratios 
$\brka$ when  only the contribution from $\paa$ is included. 
The short-dashed line corresponds to the Standard Model prediction, and 
the long-dashed line (solid line) shows the theoretical prediction for 
$F_Q=40GeV$ ($30GeV$) respectively. 

The Fig.5b shows the $\mpb$ dependence of the branching ratios $\brka$ 
when  the contributions from both $\paa$ and $\pbb$ are considered. 
The horizontal band corresponds to the experimental data (\ref{brkaexp}).
The short-dashed line shows the SM prediction, while the 
dot-dashed curve shows the
branching ratio when only the new contribution from the $\pbb$ is  taken 
into account. The long-dashed curve shows the branching ratio when 
the new contributions from both $\paa$ and $\pbb$ are 
included and assuming $\mpa=50GeV$.  The Fig.5c also show the  
mass dependence of the branching ratios $\brka$ but for $F_Q=30GeV$ 
and $\mpa=50 GeV$.

If we consider the theoretical uncertainty  of the branching ratio $\brka$ 
in the SM, 
say about $\pm 4\times 10^{-11}$\cite{buras974}, the above lower mass bounds 
will be decreased by no more than $4 GeV$. It is easy to see that the 
the uncertainty of the experimental data is still rather large and dominate
the total uncertainty.  Consequently, further reduction of the 
experimental error is very essential to constrain the Multiscale Walking 
Technicolor Model more stringently.

\section*{4. The decay $\kb$ }

 Since the rare decay $\kb$ proceeds in the SM almost entirely through 
CP violation \cite{littenberg}, it is completely dominated by 
short-distance loop effects with the top quark exchanges. The 
charm contribution can be safely neglected and there is no 
theoretical uncertainties due to $m_c, \mu_c$ and 
$\Lambda_{\overline{MS}}$ present in the decay $\ka$. At the level 
of $\brkb$ the uncertainty in the choice of $\mu_t$ is reduced from 
$\pm 10\%$ (LO) down to $\pm 1\%$ (NLO), and therefore can also 
be neglected\cite{buras974}. Consequently this decay mode
is even cleaner than $\ka$ and is very well suited for the 
probe of new physics if the experimental data can reach 
the required sensitivity. 

The effective Hamiltonian for $\kb$  is given as follows\cite{buras974}, 
\beq
{\cal H}_{eff} = \frac{G_F}{\sqrt{2}}\frac{\alpha}{2\pi \ssa}
 V^*_{ts}V_{td} X(x_t) (\bar s d)_{V-A}(\bar\nu\nu)_{V-A} + h.c.,  
\label{heffkb}
\eeq
where the functions $X(x_t)$ has been given in eq.(\ref{xxt}).

Using the effective Hamiltonian $(\ref{heffkb})$ and summing over three 
neutrino flavors one finds
\beq
\brkb = \kappa_L \cdot \left( \frac{Im \lambda_t}{\lambda^5} X(x_t) 
\right)^2 
\eeq
with $\kappa_L =1.80 \times 10^{-10} $\cite{buras974}.

In the Standard Model, using the input parameters of eq.(\ref{sip}), 
one finds 
\beq
\brkb = 2.75 \times 10^{-11}
\eeq
which is consistent with the result given in ref.\cite{buras974}.
When the contributions due to $\paa$ and $\pbb$ are 
included, the size of the 
corresponding branching ratios depends on the masses $\mpa$ and 
$\mpb$.   Using the input parameters of eq.(\ref{sip}) and assuming
$F_Q=40GeV$, $50GeV \leq \mpa \leq 400 GeV$ and 
$100GeV \leq \mpb \leq 600 GeV$, one finds 
\beq
3.42 \times 10^{-11} \leq \brkb \leq 1.31 \times 10^{-10} 
\eeq
if only the $\paa$'s contribution is included, and 
\beq
1.23 \times 10^{-11} \leq \brkb \leq 1.77 \times 10^{-9}  
\eeq
if only the $\pbb$'s contribution is included, and 
\beq
1.69 \times 10^{-10} \leq \brkb \leq 2.33 \times 10^{-9}  
\eeq
if the $\paa$'s and $\pbb$'s contributions are all included.
For the typical values of $\mpa=200 GeV$ and $\mpb=300 GeV$, one has
$\brkb(\paa)=6.03\times 10^{-11}$, $\brkb(\pbb)=2.53\times 10^{-10}$
and $\brkb(All)=3.39\times 10^{-10}$.

The Fig.6a shows the $\mpa$ dependence of the branching ratio 
$\brkb$ when  only the new contribution from $\paa$ is considered. 
The dot-dashed ( solid ) curve represents the theoretical prediction 
for $F_Q=40$ ($30$) $GeV$ respectively.
The Fig.6b shows the $\mpb$ dependence of the branching ratio $\brkb$ 
when  the contributions from both $\paa$ and $\pbb$ are considered,  
assuming $\mpa =50GeV$. The dot-dashed (solid ) curve again shows 
the theoretical prediction for $F_Q=40$ ($30$) $GeV$ respectively.

The present experimental bound on $\brkb$ from FNAL experiment E731 
\cite{weaver94} is $\brkb < 5.8 \times 10^{-5}$, which is about six orders 
of magnitude above the SM expectation and about four orders of magnitude 
above the theoretical prediction when the maximum new contributions from  
the charged technipions are included.
There is obviously a long way to go for the 
forthcoming or planed experiments \cite{ags2000,arisaka91} 
to measure this gold-plated decay mode with
enough sensitivity to probe the effects of new physics. 

\section*{5. The decay $\kc$ }

For the decay $\kc$, the situation is more complicated because of the
presence of long-distance contributions from the two-photon 
intermediated state which are difficult to calculate reliably 
\cite{eeg96}. But one important advantage is the availability of the 
experimental data with good sensitivity\cite{epjc98}. 
In this paper we only consider the new physics effects 
to the short distance part $\kcsd$.  

In the Standard Model, 
the effective Hamiltonian for $\kc$  are now available at the NLO level
\cite{buras974}, 
\beq
{\cal H}_{eff} = - \frac{G_F}{\sqrt{2}}\frac{\alpha}{2\pi \ssa}
 \left[ V^*_{cs}V_{cd} Y_{NL} + V^*_{ts}V_{td} Y(x_t) \right] 
(\overline{s}d)_{V-A}(\overline{\mu}\mu)_{V-A} + h.c. 
\label{heffkc}
\eeq
where the functions $Y(x_t)$ and $Y_{NL}$ have been given in 
eqs.(\ref{yxt},\ref{ynl}).

Using the effective Hamiltonian $(\ref{heffkc})$ and relating 
$<0|(\overline{s}d)_{V-A}|K_L>$ to $Br(K^+ \to \mu^+ \nu)$ one finds
\cite{buras94,buras961} 
\beq
\brkcsd = \kappa_{\mu} \cdot \left[ 
\frac{Re \lambda_c}{\lambda}P_0(Y)^{SM} + 
\frac{Re \lambda_t}{\lambda^5}Y(x_t) \right]^2
\eeq
with  $\kappa_{\mu} =1.68 \times 10^{-9} $\cite{buras974}.

Within the SM, using the input parameters of eq.(\ref{sip}), one finds
\beq
\brkcsd = 1.25 \times 10^{-9}
\eeq
which is consistent with the result as given in ref.\cite{buras974}. 
When the long-distance part is also included\cite{buras974} one finds, 
\beq 
\brkc_{TH}= (6.81 \pm 0.32)\times 10^{-9}
\label{brkcth}
\eeq
which is basically consistent with the data
\cite{epjc98},
\beq
\brkc = (7.2 \pm 0.5)\times 10^{-9} 
\label{brkcexp}
\eeq
the error of the data will be reduced to about $\pm 1\%$ at 
BNL in the next years.

When the new contributions due to $\paa$ and $\pbb$ are included, 
one finds
\beq
\brkcsd = \kappa_{\mu} \cdot \left[ 
\frac{Re \lambda_c}{\lambda}P_0(Y) + 
\frac{Re \lambda_t}{\lambda^5}Y(x_t,y_t,z_t) \right]^2
\eeq
where $\kappa_{\mu} =1.68 \times 10^{-9} $\cite{buras974}.
By using the input parameters of eq.(\ref{sip}) and assuming $F_Q=40GeV$, 
$50GeV \leq \mpa \leq 400 GeV$ and $100GeV \leq \mpb \leq 600 GeV$, 
one has 
\beq
1.71\times 10^{-9} \leq \brkcsd \leq 7.55 \times 10^{-9} 
\eeq
if only the $\paa$'s contribution is included, and 
\beq
0.99 \times 10^{-9} \leq \brkcsd \leq 1.19 \times 10^{-7}  
\eeq
if only the $\pbb$'s contribution is included, and 
\beq
1.42\times 10^{-9} \leq \brkcsd \leq 1.57 \times 10^{-7}  
\eeq
if the $\paa$'s and $\pbb$'s contribution are all included.
For the typical values of $\mpa=200 GeV$ and $\mpb=300 GeV$, one finds
$\brkcsd(\paa)=3.24\times 10^{-9}$, $\brkcsd(\pbb)=1.72\times 10^{-8}$
and $\brkcsd (All)=2.34\times 10^{-8}$.

The Fig.7a shows the $\mpa$ dependence of the branching ratio 
$\brkcsd$ when  only the extra contribution from $\paa$ is included, 
where the solid ( dot-dashed ) curve corresponds to the theoretical 
predictions for $\brkcsd$ with the inclusion of the contribution due to
$\paa$ for $F_Q=30, 40GeV$ respectively. The Fig.7b shows the $\mpb$ 
dependence of the branching ratio $\brkcsd$ 
when  the contributions from both $\paa$ and $\pbb$ are considered. 
The short-dashed line shows the SM prediction $\brkcsd = 1.25 \times 
10^{-9}$, while the dot-dashed line shows the branching ratio 
$\brkcsd = 7.55\times 10^{-9}$ for $\mpa=50GeV$. 
The long-dashed curve shows the
branching ratio when only the extra contribution from the $\pbb$ is 
included. The solid curve corresponds to the branching ratio 
when all new contributions are taken into account.

The situation for the decay $\kc$ is rather subtle because of the 
involvement
of the long-distance part. Firstly, the experimental data is accurate 
and  basically consistent with the current  theoretical predictions for the 
decay $\kc$ in the SM. Secondly, the calculation
for the  short-distance part is rather reliable. And finally the size of the
new physics contributions strongly depend on the masses of new particles
as shown in Fig.7.  It seems that
this decay mode should  be very helpful for us to test the SM and to 
probe the effects of the new physics beyond the SM, or at least 
to put some limits on 
the masses of new particles. But it is very difficult to calculate the 
long-distance contribution reliably, 
the current result is only an estimation based on some general
assumptions and inevitably has large uncertainty. This fact makes it 
difficult to get a reliable theoretical prediction for the decay $\kc$ 
at present. 

As an estimation, we at first conservatively assume that 
the uncertainty 
of the current theoretical prediction for the decay $\kc$ is 
two times larger than that given in eq.(\ref{brkcth}), i.e.,  
\beq
\brkc_{TH} = (6.81 \pm 0.96)\times 10^{-9}
\label{brkcthl}
\eeq
and to see if we can find any bounds on the masses of the $\paa$ and 
$\pbb$ by comparing the theoretical prediction (\ref{brkcthl}) with 
the experimental data (\ref{brkcexp}).

Fig.8a is the plot of the branching ratio $\brkc$ as a function of the mass 
$\mpa$ assuming $F_Q=40GeV$. The three curves are the theoretical 
predictions with the inclusion of the new contribution from $\paa$ only, 
and  the horizontal band shows the experimental data (\ref{brkcexp}). 
One can read the lower bound on $\mpa$ from the Fig.8a:
\beq
\mpa \geq 210 GeV\label{bound1}
\eeq
when we neglect the contributions to the rare decay $\kc$ from the $\pbb$. 
For $F_Q=30GeV$, the corresponding lower bound is $\mpa \geq 313 
GeV$. If we treat the theoretical prediction $\brkc_{TH} =6.81 \pm 0.32$ as 
a reliable one the corresponding lower mass bounds are
$\mpa \geq 262,$  or, $355 GeV$ for $F_Q=40, 30GeV$ respectively.
For $F_Q=40GeV$ and $\mpb \leq 490 GeV$ the whole parameter space for 
$\mpa$ is excluded completely. 

From Fig.8b  one can read the constraints on the $\mpb$.
The horizontal band corresponds to the data, while the three curves 
are the theoretical predictions with the inclusion
of the contributions from the $\pbb$ only. By comparing the
theoretical predictions ( with the enlarged theoretical uncertainty 
$\pm 0.96$)  with the data one finds the lower bounds on
$\mpb$:
\beq  
\mpb \geq  \left \{ \begin{array}{ll} 
490 GeV &  {\rm for} \ \ F_Q=40 \ \ {\rm GeV}  \\
527 GeV &  {\rm for} \ \ F_Q=30 \ \ {\rm GeV} 
\end{array} \right. 
\label{bound8a}
\eeq
If we take the theoretical uncertainty $\pm 0.32$ 
as a reliable one, the above lower bounds on $\mpb$ will be increased by 
about $20 GeV$. 

Fig.8c is the plot of the branching ratio $\brkc$ as a function of the mass 
$\mpb$ assuming $\mpa=210GeV$ and $F_Q=40GeV$.  The three curves are the 
theoretical predictions with the inclusion
of the contributions from the $\paa$ and $\pbb$. By comparing the
theoretical predictions ( with the enlarged theoretical uncertainty 
$\pm 0.96$)  with the data one finds the lower bounds on
$\mpb$:
\beq  
\mpb \geq  \left \{ \begin{array}{ll} 
580 GeV &  {\rm for} \ \ F_Q=40 \ \ {\rm GeV}  \\
630 GeV &  {\rm for} \ \ F_Q=30 \ \ {\rm GeV} 
\end{array} \right. 
\label{bound8b}
\eeq
If we take the theoretical uncertainty $\pm 0.32$ 
as a reliable one, the above lower bounds on $\mpb$ will be increased by 
about $20 GeV$. 

One can see from (\ref{bound1},\ref{bound8a},
\ref{bound8b}) the lower bounds on $\mpa$ and $\mpb$ are rather stringent.
Although the theoretical uncertainty for the long-distance part of the 
branching ratio $\brkc$ is still large, but the current experimental data
leads to a meaningful and stringent constraints on the mass spectrum of 
unit-charged technipions and consequently on the multiscale walking 
technicolor model itself. 

In ref.\cite{lane91}, the authors 
constructed a specific multiscale walking technicolor model and calculated 
the dijet and technipion production rates at the hadron colliders by using 
two sets of input mass parameters (\ref{seta},\ref{setb}). 
But the assumed mass ranges of the $\paa$ and $\pbb$ are clearly excluded 
by the constraints from the rare K-decay process $\kc$ 
as given in (\ref{bound1}, \ref{bound8a}, \ref{bound8b}). 
Although the detailed study about the specific 
model constructed in ref.\cite{lane91} is clearly beyond the scope
of this paper, but the assumed mass spectrums for unit-charged 
technipions as given in  ref.\cite{lane91} are excluded by the data,  
according to our calculations.

\section*{6. Conclusion and discussions}
 
In this paper we calculate the $Z^0-$penguin contributions to the rare 
FCNC K-decays $\ka$, $\kb$ and $\kc$ 
from the  unit-charged technipions $\paa$ and $\pbb$ 
appeared in the MWTCM \cite{lane91}. 

We firstly evaluate the new $Z^0$-penguin diagrams induced by the 
$\paa$ and $\pbb$, and extract the  finite functions $C_0^{New}(y_j)$ 
, $C_0^{New}(z_j)$, $C_{NL}(\paa)$ and $C_{NL}(\pbb)$ which govern  
the new contributions to the decay in question and plays the same rule 
as the functions $C_0(x_i)$ and $C_{NL}$ in ref.\cite{buras974} for 
the study of rare K-decays. 
The color-octet $\pbb$ does not contribute to the decay 
through the box-diagrams, while the tiny box-diagram contributions 
from $\paa$ can be neglected safely.
The charged technipions contribute to the branching ratios of the 
rare K-decays through the functions $C_0^{New}$ and 
$C_{NL}^{New}$ by a proper linear combination with their 
Standard Model counterparts $C_0(x_t)$ and $C_{NL}$.

The size of the new contributions generally depends on the value of the 
technipion decay constant $F_Q$  and the mass spectrum of the 
charged technipions, and the color-octet $\pbb$ dominant  in a large part 
of the parameter space. At the level of the corresponding 
branching ratios, the maximum enhancement due to $\paa$ is about 
one order of magnitude. While the maximum enhancement  
due to $\pbb$ can be as large as two orders. So strong 
enhancements to the relevant branching ratios of $\brka$ and $\brkc$
make it possible to put some constraints on the mass spectrum 
of charged technipions by comparing the theoretical predictions with 
the experimental data available. 

For the decay $\ka$, as illustrated in Figs.(5a, 5b, 5c),  
there is no independent constraint on $\mpa$ at present, but further 
refinement of the data may put some constraints on $\mpa$ in the near 
future. For the color-octet technipion $\pbb$, the typical constraints 
are $\mpb \geq 228GeV$ assuming $F_Q=40GeV$ and $\mpa=200GeV$, and 
$\mpb \geq 341GeV$ assuming $F_Q=30GeV$ and $\mpa=200GeV$. 

For the decay $\kb$, as shown in Figs.(6a, 6b), 
no  constraint on both $\mpa$ and $\mpb$ can be derived now because 
of the low sensitivity of the available data.

For the decay $\kc$, the situation is rather subtle because of the 
involvement of the long-distance part. Our attempt to 
constrain the new physics models is hampered to some degree by the large  
uncertainty of the long-distance piece of the branching ratio $\brkc$. 
Fortunately, thanks to the accurate experimental data published by the E787 
collaboration,  rather strong constraints have been obtained even if we use 
the enlarged uncertainty of $\brkc_{TH}$. 

For the color-singlet $\paa$, the lower mass bound is $\mpa \geq 210 GeV$ 
if we neglect the new contribution to the decay $\kcsd$ from the 
color-octet $\pbb$. For $F_Q=30GeV$, the corresponding lower bound 
is $\mpa \geq 313 GeV$.
If we treat the theoretical prediction $\brkc_{TH} =6.81 \pm 0.32$ as 
a reliable one the corresponding lower mass bounds are
$\mpa \geq 262,$  or, $355 GeV$ for $F_Q=40, 30GeV$ respectively.
For $F_Q=40GeV$ and $\mpb \leq 490 GeV$, the whole assumed parameter 
space for $\paa$, $50 GeV \leq \mpa \leq 400GeV$, is excluded completely 
by the data. 

For color-octet technipion $\pbb$, the lower bounds on $\mpb$ are much 
stronger than that on $\mpa$. If we neglect the $\paa$'s contributions to
the branching ratio $\brkcsd$ and use the enlarged theoretical uncertainty 
$\delta \brkc_{TH} =0.96$, the lower bounds on $\mpb$ are 
$\mpb \geq  490 GeV$ ($527GeV$) for $ F_Q=40GeV$ ($30GeV$). 
The above lower bounds on $\mpb$ will be 
increased by about $20 GeV$, If we take the theoretical uncertainty 
$\pm 0.32$ as a reliable one. 

If we take into account the contributions due to the $\paa$ (  
assuming $\mpa=210GeV$ ) and use the enlarged theoretical uncertainty 
$\delta \brkc_{TH} =0.96$,  the lower bounds on $\mpb$ are 
$\mpb \geq  580 GeV$ ($630GeV$) for $ F_Q=40GeV$ ($30GeV$). 
The above lower bounds on $\mpb$ will be 
increased again by about $20 GeV$, If we take the theoretical uncertainty 
$\pm 0.32$ as a reliable one.

For intrinsic and technical reasons, it is very difficult to calculate 
the strong ETC and walking technicolor interactions reliably. But one can 
use the currently known knowledge  to  make primary
estimations about the possible contributions to various physical processes 
from the new particles appeared in the MWTCM, and in turn to test 
the model itself or constrain the parameter space of the model. In 
ref.\cite{lu97a}, the authors examined the corrections to the branching 
ratio $R_b=\Gamma(Z\to b \bar b)/\Gamma(Z \to hadron)$ due to the exchanges
of the ETC gauge bosons and found that 
the new contribution is too large to be consistent with the LEP data in 
most of the parameter space in the MWTCM\cite{lu97a}. 
In ref.\cite{lu97b}, the authors estimated the corrections to the rare 
decay $b \to s \gamma$ in the MWTCM, and found that 
the whole range of $\mpb \leq 600 GeV$ is excluded by the CLEO data of 
$Br(B \to X_s \gamma) = (2.32 \pm 0.57 \pm 0.35) \times 
10^{-4}$\cite{lu97b}.
In this paper, we studied the new contributions to the rare K-decays 
from the unit-charged technipions in the framework of the MWTCM. 
The resulted constraints on the $\mpa$ and $\mpb$ from the data  
(\ref{brkcexp}) are rather stringent.
The main cause which leads to the above constraints is the smallness 
of the $F_Q$ (which is clearly a basic feature of the MWTCM ).
If we treat above results seriously, one conclusion is inevitable: 
the smallness of $F_Q$ is disfavored by the $\brkc$ and $R_b$ data, and 
the assumed mass parameter
space for the  $\paa$ and $\pbb$ as given in ref.\cite{lane91} is excluded 
by the data (\ref{brkcexp}) as well as the CLEO data for rare decay 
$b \to s \gamma$, and therefore the multiscale walking 
technicolor model itself as constructed in ref.\cite{lane91} is 
strongly disfavored by the data. One way out is to modify the multiscale 
walking technicolor model by introducing the Topcolor interaction 
\cite{hill95} into the model \cite{xiao991}.

\vspace{.5cm}

\noindent
{\bf Note added:} In the calculation of the branching ratios of the decay 
$\kc$, we neglected the dispersive part  $A_{LD}$ of the long-distance 
contribution. In fact the measured rate $\brkc$ is almost 
saturated by the absorptive contribution, leaving only a small room 
for the coherent sum of the long- and short-distance dispersive contribution.
Therefore, the magnitude of the total real part $Re[A]$ must be 
relatively small compared with the absorptive part. 
Such a small total dispersive 
amplitude can be realized either when the $A_{SD}$  and $A_{LD}$ parts 
are both small (this is the case assumed in this paper) or by partial 
cancellation between these two parts as being considered in 
ref.\cite{xiao991}. Even if we take into account the effects of the term 
$A_{LD}$, the conclusion of this paper still remain unchanged: 
the new contribution to the ratio $\brkc$ in the multiscale walking 
technicolor model is too large to be consistent with the data.

\section*{ACKNOWLEDGMENT}

Z. Xiao would like to thank R.G. Roberts and all friends in the  
Theory Group at the Rutherford-Appleton 
Laboratory for their hospitality and support. This work was partly done 
during my stay at RAL as a visitor. This work is  supported  by the 
National Natural Science Foundation of China under Grant No.19575015 
and  by the Sino-British Friendship Scholarship Scheme.

\section*{Appendix A}

In this Appendix, we present the explicit expressions for 
the functions $C_0(x_t)$, $B_0(x_t)$, $X_1(x_t)$, $Y_1(x_t)$, 
$C_{NL}$, $B_{NL}^{1/2}$,  $B_{NL}^{-1/2}$, $C_{NL}(\paa)$ and 
$C_{NL}(\pbb)$. One can also find the  
expressions for the first seven functions in ref.\cite{buras961}. 

The functions of $C_0(x_t)$ and $B_0(x_t)$ govern the leading top 
quark contributions through the $Z^0$-penguin and W-box diagrams 
in the SM, while the functions $X_1(x_t)$ and $Y_1(x_t)$ describe
the NLO QCD corrections,
\beq
B_0(x_t)&=& \frac{1}{4} \left[ \frac{x_t}{1-x_t} 
+ \frac{x_t \ln[x_t]}{(x_t-1)^2} \right] \\
C_0(x_t)&=& \frac{x_t}{8} \left[ \frac{x_t-6}{x_t-1} 
+ \frac{3x_t + 2}{(x_t-1)^2}\, \ln[x_t] \right]\\
X_1(x_t) &=& -\frac{23x_t + 5x_t^2 -4x_t^3}{3(1-x_t)^2} 
+ \frac{x_t -11x_t^2 +x_t^3 + x_t^4}{(1-x_t)^3}\ln[x_t] \nonumber \\ 
&\ \ \ \ \ + & \frac{8x_t +4x_t^2+x_t^3 - x_t^4}{2(1-x_t)^3}\ln^2[x_t]
- \frac{4x_t -x_t^3}{(1-x_t)^2}L_2(1-x_t) \nonumber \\
&\ \ \ \ \ +& 8x_t\frac{\partial X_0(x_t)}{\partial x_t}\ln[x_{\mu}] \\
Y_1(x_t) &=& -\frac{4x_t + 16x_t^2 +4x_t^3}{3(1-x_t)^2} 
- \frac{4x_t -10x_t^2 -x_t^3 - x_t^4}{(1-x_t)^3}\ln[x_t] \nonumber \\ 
&\ \ \ \ \ + & \frac{2x_t -4x_t^2+x_t^3 - x_t^4}{2(1-x_t)^3}\ln^2[x_t]
- \frac{2x_t +x_t^3}{(1-x_t)^2}L_2(1-x_t) \nonumber \\
&\ \ \ \ \ +& 8x_t\frac{\partial Y_0(x_t)}{\partial x_t}\ln[x_{\mu}] 
\eeq
where $x_t=m_t^2/m_W^2$,  $x_{\mu}=\mu^2/M_W^2$ with 
$\mu = {\cal O}(m_t)$ and 
\beq
L_2(1-x_t)= \int_1^{x_t}\; dy\, \frac{ln[y]}{1-y}.
\eeq

For the charm sector, the $C_{NL}$ is the $Z^0$-penguin part and 
the $B_{NL}^{1/2}$ ($B_{NL}^{-1/2}$) is the box contribution, 
relevant for the case of final state leptons with $T_3=1/2$ ($T_3=-1/2$): 
\beq
C_{NL} &=& \frac{x(m)}{32}\,K_c^{24/25}\,\left[ \left(
\frac{48}{7}K_+ +\frac{24}{11}K_-  -\frac{ 696}{77}K_{k33} 
\right)\left(\frac{4\pi}{\alpha_s(\mu)} + \frac{15212}{1875}( 1-K_c^{-1}) 
\right) \right. \nonumber \\ 
 & \ \ \ +&  \left( 1-\ln \frac{\mu^2}{m^2} \right) (16K_+ -8K_-) 
  - \frac{1176244}{13125}K_+ -\frac{2302}{6875}K_-  +  
\frac{3529184}{48125}K_{33} \nonumber \\
 &\ \ \ +& \left. K\, \left( \frac{56248}{4375}K_+ -\frac{81448}{6875}K_- 
+\frac{4563698}{144375}K_{33} \right) \right] 
\label{cnl}
\eeq
where
\beq 
K = \frac{\alpha_s(M_W)}{\alpha_s(\mu)}, \ \ 
K_c = \frac{\alpha_s(\mu)}{\alpha_s(m)}, \ \ 
K_+ = K^{6/25},\ \  K_- = K^{-12/25},\ \  K_{33} = K^{-1/25} 
\eeq
and 
\beq
B_{NL}^{1/2}&=& \frac{x(m)}{4}\,K_c^{24/25}\,\left[3(1-K_2)\left( 
\frac{4\pi}{\alpha_s(\mu)}
   + \frac{15212}{1875}(1-K_c^{-1}) \right) \right.\nonumber \\
&\ \ \ - & \left. \ln\frac{\mu^2}{m^2}
   - \frac{r\ln r}{1-r} -\frac{305}{12} + \frac{15212}{625}K_2 
   + \frac{15581}{7500} K\, K_2 \right] \\
B_{NL}^{-1/2}&=& \frac{x(m)}{4}\,K_c^{24/25}\,\left[
3(1-K_2)\left( \frac{4\pi}{\alpha_s(\mu)}
   + \frac{15212}{1875}(1-K_c^{-1}) \right) \right. \nonumber \\
&\ \ \ - & \left. 
\ln\frac{\mu^2}{m^2}-\frac{329}{12} + \frac{15212}{625}K_2 
   + \frac{30581}{7500} K\, K_2 \right] 
\eeq
here $K_2=K_{33}$, $m= m_c$, $\mu = {\cal O}(m_c)$, $x(m)=m_c^2/M_W^2$,  
$r=m_l^2/m_c^2(\mu)$ and $m_l$ is the lepton mass. 

For the charm sector, the functions of $C_{NL}(\paa)$ and $C_{NL}(\pbb)$ 
describe the contributions from the $\paa$ and $\pbb$,
\beq
C_{NL}(\paa) &=& a_1\,K_c^{24/25}\,\left[ \left(
\frac{48}{7}K_+^{\paa} +\frac{24}{11}K_-^{\paa}  
-\frac{ 696}{77}K_{k33}^{\paa} 
\right)\left(\frac{4\pi}{\alpha_s(\mu)} + \frac{15212}{1875}( 1-K_c^{-1}) 
\right) \right. \nonumber \\ 
& + & \left( 1-\ln \frac{\mu^2}{m^2} \right) (16K_+^{\paa} 
-8K_-^{\paa}) 
  - \frac{1176244}{13125}K_+^{\paa} -\frac{2302}{6875}K_-^{\paa}  +  
\frac{3529184}{48125}K_{33}^{\paa} \nonumber \\
& +& \left. K_{\paa}\, \left( \frac{56248}{4375}K_+^{\paa} 
-\frac{81448}{6875}K_-^{\paa} 
+\frac{4563698}{144375}K_{33}^{\paa} \right) \right] 
\label{cnlb}
\eeq
with
\beq 
a_1& = &\frac{m_c^2}{768\, \sqrt{2}\,F_Q^2\,G_F\,M_W^2},  \ \  
K_{\paa} = \frac{\alpha_s(\mpa)}{\alpha_s(\mu)}, \ \  
K_c = \frac{\alpha_s(\mu)}{\alpha_s(m)}, \nonumber \\ 
K_+^{\paa} &=& (K_{\paa})^{6/25},\ \  K_-^{\paa} = (K_{\paa})^{-12/25},\ \  
K_{33}^{\paa} = (K_{\paa})^{-1/25} 
\eeq
and 
\beq
C_{NL}(\pbb) &=& a_8 \,
K_c^{24/25}\,\left[ \left( \frac{48}{7}K_+^{\pbb} +\frac{24}{11}K_-^{\pbb} 
 -\frac{ 696}{77}K_{k33}^{\pbb} 
\right)\left(\frac{4\pi}{\alpha_s(\mu)} + \frac{15212}{1875}( 1-K_c^{-1}) 
\right) \right. \nonumber \\ 
& + & \left( 1-\ln \frac{\mu^2}{m^2} \right) (16K_+^{\pbb} -
8K_-^{\pbb}) 
  - \frac{1176244}{13125}K_+^{\pbb} -\frac{2302}{6875}K_-^{\pbb}  +  
\frac{3529184}{48125}K_{33}^{\pbb} \nonumber \\
& +& \left. K_{\pbb}\, \left( \frac{56248}{4375}K_+^{\pbb} 
-\frac{81448}{6875}K_-^{\pbb} 
+\frac{4563698}{144375}K_{33}^{\pbb} \right) \right] 
\label{cnlc}
\eeq
with
\beq 
a_8 & = &\frac{m_c^2}{96\, \sqrt{2}\,F_Q^2\,G_F\,M_W^2},  \ \  
K_{\pbb} = \frac{\alpha_s(\mpb)}{\alpha_s(\mu)}, \ \ 
K_c = \frac{\alpha_s(\mu)}{\alpha_s(m)}, \nonumber \\ 
K_+^{\pbb} &=& (K_{\pbb})^{6/25},\ \  K_-^{\pbb} = (K_{\pbb})^{-12/25},\ \  
K_{33}^{\pbb} = (K_{\pbb})^{-1/25}. 
\eeq

\vspace{0.5cm}

\vspace{.5cm}
\begin{center}
{\bf Figure Captions}
\end{center}
\begin{description}

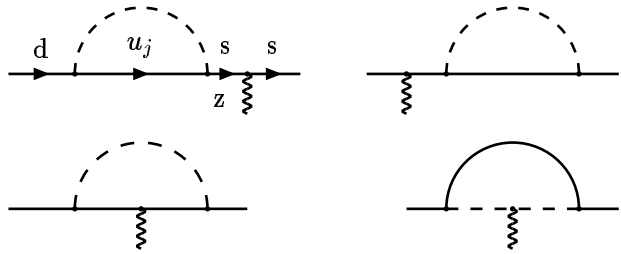
\item[Fig.1:] The new $Z^0-$penguin diagrams contributing to the induced 
$\dsz$ vertex from the internal exchanges of the technipion 
$\paa$ and $\pbb$. The dashed lines are $\paa$ and $\pbb$ lines and the 
$u_j$ stands for the quarks $(u, c, t)$.

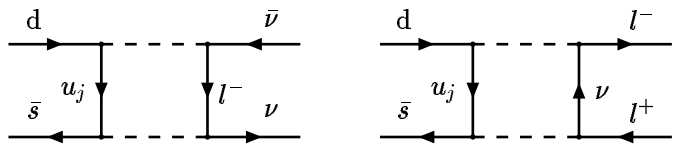
\item[Fig.2:] The new box diagrams contributing to the studied processes 
by internal exchanges of color-singlet $\paa$. 

\item[Fig.3:] The Figs.(3a, 3b) are the plots of the functions 
$X(x_t)$ and $Y(x_t)$ vs the mass $\mpb$.
 For more details see the text.

\item[Fig.4:] The Figs.(4a, 4b)  are the plots of the functions 
$P_0(X)$ and $P_0(Y)$ vs the mass $\mpb$. For more details see the text.

\item[Fig.5:] The Fig.5a is the plot of the branching ratio $\brka$ vs 
the mass $\mpa$ for $F_Q=30, 40GeV$ respectively. 
The Figs.(5b, 5c)  are the plots of the branching ratio 
 $\brka$ vs the mass $\mpb$ for $F_Q=30, 40GeV$ respectively.
 For more details see the text.

\item[Fig.6:] The Figs.(6a, 6b)  are the plots of the branching ratio 
 $\brkb$ vs the mass $\mpa$ and $\mpb$ and assuming 
$F_Q=30, 40GeV$ respectively. For more details see the text.

\item[Fig.7:] The Figs.(7a, 7b)  are the plots of the branching ratio 
 $\brkcsd$ vs the mass $\mpa$ and $\mpb$ respectively.
 For more details see the text.

\item[Fig.8:] The Figs.(8a,8b) show the lower bounds 
on the mass of the $\paa$ and $\pbb$ assuming $F_Q= 40 GeV$. 
The horizontal band corresponds  to the current experimental data,
the three curves are the theoretical predictions when  we 
use the enlarged uncertainty of $Br(K_L \to \mu^+ \mu^-)_{TH}$.
The Fig.8c shows the lower bounds on the mass of the $\pbb$ if we 
use $\pm 0.32$ instead of the enlarged $\pm 0.96$ as the uncertainty 
of the theoretical prediction. One can read the lower bounds on the 
masses of the $\paa$ and $\pbb$. 

\end{description}

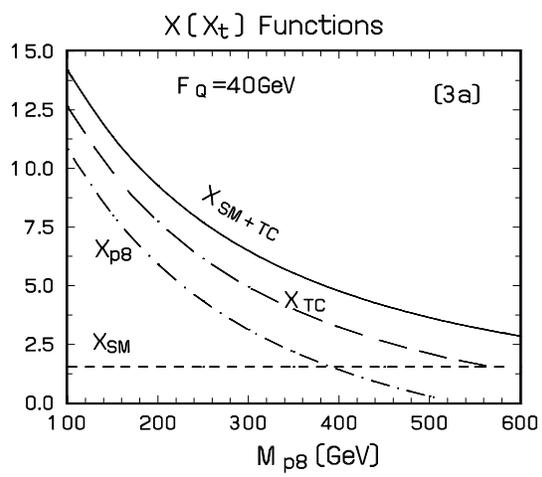
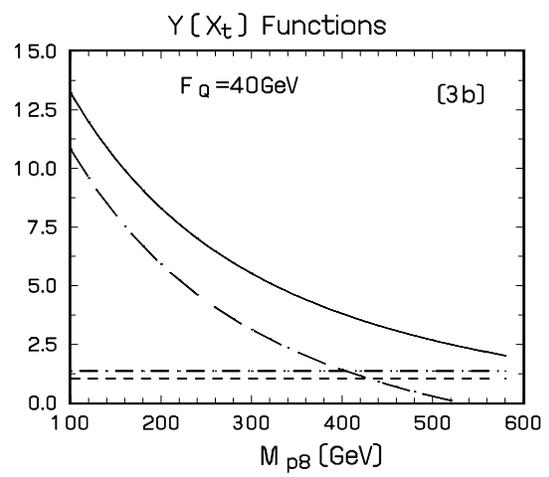
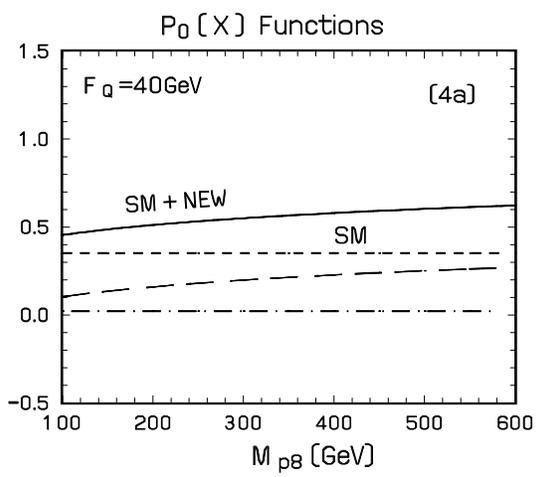
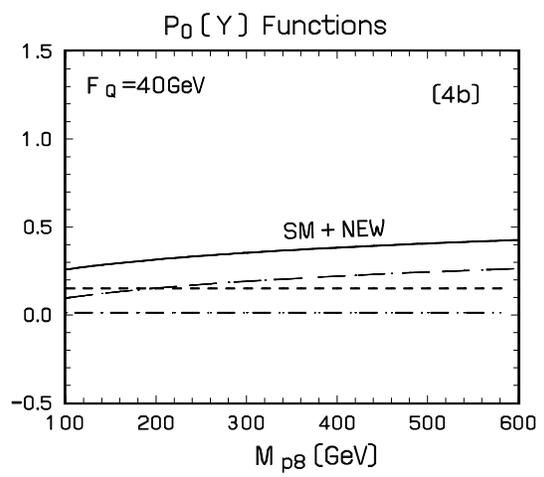
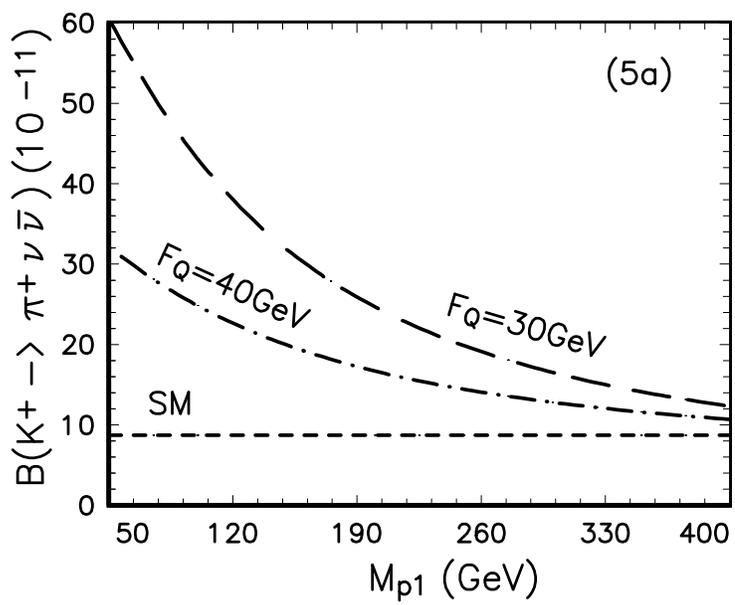
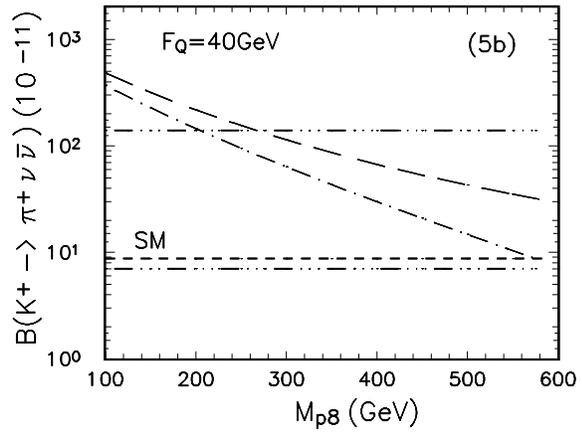
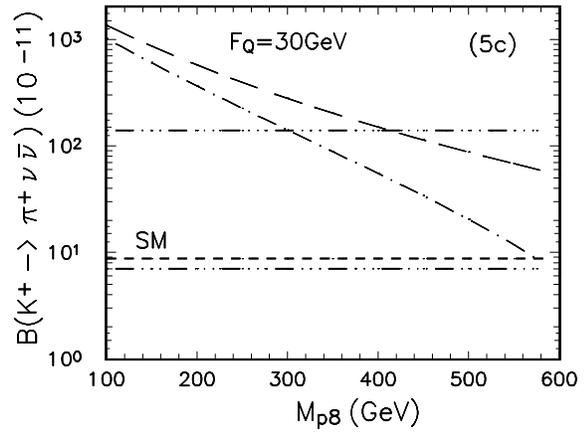
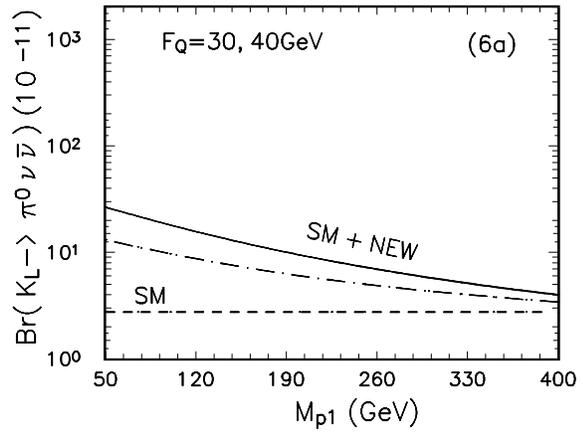
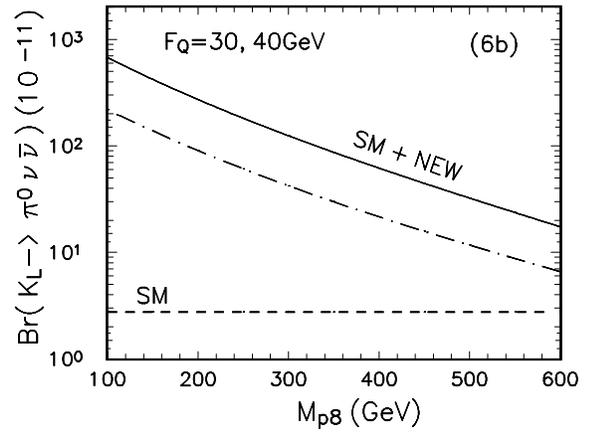
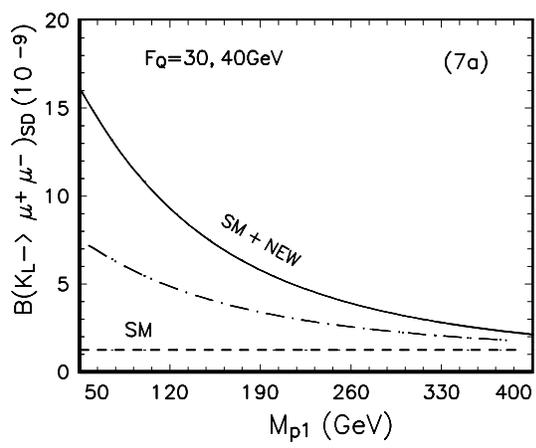
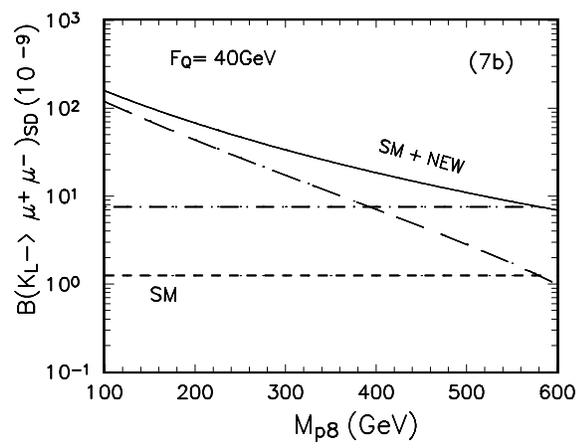
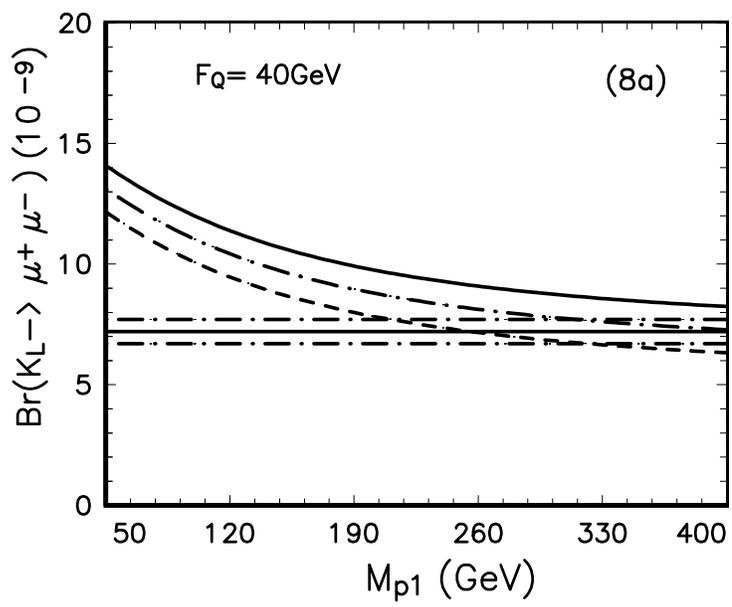
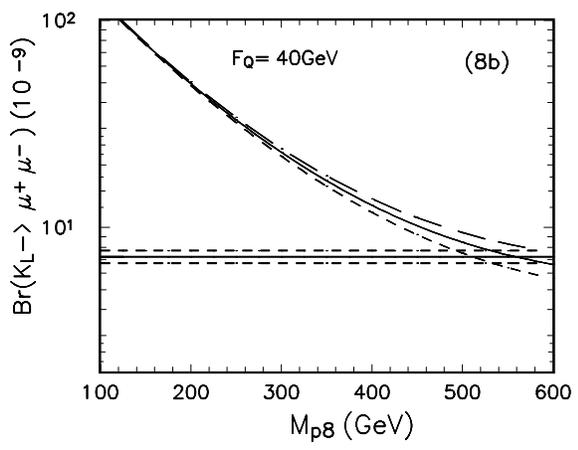
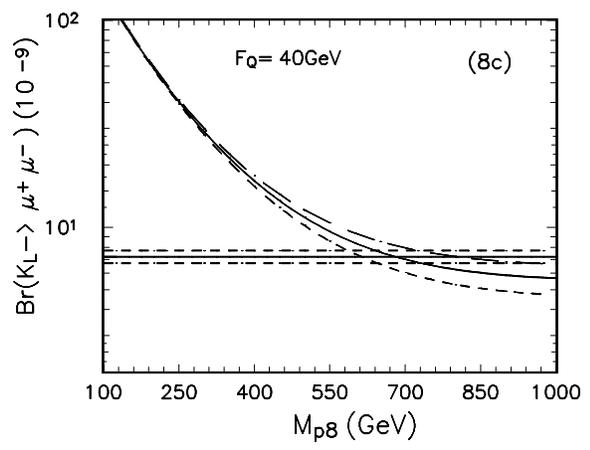
\end{document}